\documentclass[noacm, sigconf]{acmart}

\usepackage{stfloats}

\usepackage{amsmath,amsfonts}
\usepackage{multirow}

\usepackage{listings}
\usepackage{graphicx}
\usepackage{textcomp}

\usepackage{xcolor}

\usepackage{caption}
\usepackage{subcaption}

\usepackage{enumitem}
\usepackage{tabularx}

\usepackage{pbalance}

\begin{document}

\title[Register Dispersion: Reducing the 
Footprint of the Vector Register File]
{Register Dispersion: Reducing the 
Footprint of the Vector Register File in Vector Engines of Low-Cost RISC-V CPUs}

\author{Vasileios Titopoulos}
\affiliation{%
  \institution{Electrical and Computer Engineering Democritus University of Thrace}
  \city{Xanthi}
  \country{Greece}
}
\author{George Alexakis}
\affiliation{%
  \institution{Electrical and Computer Engineering Democritus University of Thrace}
  \city{Xanthi}
  \country{Greece}
}
\author{Kosmas Alexandridis}
\affiliation{%
  \institution{Electrical and Computer Engineering Democritus University of Thrace}
  \city{Xanthi}
  \country{Greece}
}
\author{Chrysostomos Nicopoulos}
\affiliation{%
  \institution{Electrical and Computer Engineering University of Cyprus}
  \city{Nicosia}
  \country{Cyprus}
}
\author{Giorgos Dimitrakopoulos}
\affiliation{%
  \institution{Electrical and Computer Engineering Democritus University of Thrace}
  \city{Xanthi}
  \country{Greece}
}

\begin{abstract}

The deployment of Machine Learning (ML) applications at the edge on resource-constrained devices has accentuated the need for 
efficient ML processing on low-cost processors. While traditional CPUs provide programming flexibility, their general-purpose architecture often lacks the throughput required for complex ML models. The augmentation of a RISC-V processor with a vector unit can provide substantial data-level parallelism. However, increasing the data-level parallelism supported by vector processing would make the Vector Register File (VRF) a major area consumer in ultra low-cost processors, since 32 vector registers are required for RISC-V Vector ISA compliance. This work leverages the insight that many ML vectorized kernels require a small number of active vector registers, and proposes the use of a physically smaller VRF that dynamically caches only the vector registers currently accessed by the application. This approach, called \textit{Register Dispersion}, maps the architectural vector registers to a \textit{smaller} set of physical registers. The proposed ISA-compliant VRF is significantly smaller than a full-size VRF and operates like a conventional cache, i.e., it only stores the most recently accessed vector registers. Essential registers remain readily accessible within the compact VRF, while the others are offloaded to the cache/memory sub-system. The compact VRF design is demonstrated to yield substantial area and power savings, as compared to using a full VRF, with no or minimal impact on performance. This effective trade-off renders the inclusion of vector units in low-cost processors feasible and practical.
\end{abstract}




\copyrightyear{2025}
\acmYear{2025}
\setcopyright{none}
% \setcctype{by}
\acmConference[CF '25]{22nd ACM International Conference on Computing Frontiers}{May 28--30, 2025}{Cagliari, Italy}
\acmBooktitle{22nd ACM International Conference on Computing Frontiers (CF '25), May 28--30, 2025, Cagliari, Italy}\acmDOI{10.1145/3719276.3725181}
\acmISBN{}
\renewcommand{\shortauthors}{V. Titopoulos et al.}

% \end{frontmatter}
\maketitle

\section{Introduction}
\label{sec:introduction}

The widespread proliferation of Artificial Intelligence (AI) and Machine Learning (ML) in a multitude of application domains is advancing at rapid pace. In addition to extensive cloud-based deployment, ML applications are increasingly being utilized at the edge~\cite{ML_on_edge}, and often on very resource-constrained devices that impose stringent hardware area and power requirements. Nevertheless, the required performance when executing the ML models remains high enough to guarantee a %seamless and 
high-quality experience for the end user. This conflicting demand for high application performance on ultra low-cost hardware necessitates novel architectural solutions that can effectively reconcile the contrasting requirements.

Vector processing~\cite{vector-processors,bigvlittle,vec_for_emb} has re-emerged as a highly effective and efficient computational paradigm to accelerate data-level parallel applications. By utilizing wide vector registers and multiple execution units (aka \textit{vector lanes}), vector processors can achieve very high computational throughput by simultaneously operating, in parallel, on multiple 
data elements~\cite{xuantie, bigvlittle, ara, Riscv2, vitruvius}. 

Several modern microprocessors enhance their processing cores with dedicated vector units to effectively exploit Data-Level Parallelism (DLP). These Vector Processing Units (VPU) are accessed through various vector instructions that have been added to the native Instruction Set Architecture (ISA) in the form of ISA extensions. One such example is the RISC-V ``V'' Vector (RVV) Extension that adds vector computation capabilities to the RISC-V ISA. In this context, several vector processors~\cite{ara,Riscv2, ara2,araxl,vitruvius,vicuna,saturn} have been proposed for RVV.

Compatibility with the RVV ISA specification necessitates a 32-entry Vector Register File (VRF). Traditional vector architectures include a \emph{dedicated} VRF for achieving sufficient performance.
A dedicated VRF, particularly in architectures with long vector lengths (exceeding 256 bits)~\cite{araxl}, or even in those with shorter vectors~\cite{saturn}, significantly impacts processor area~\cite{vec_lim,Spatz,araxl,Riscv2}. While acceptable in high-end processors, this area overhead can be prohibitive for resource-constrained edge devices. Other low-cost alternatives operate without a separate VRF~\cite{Ahromma,sparrow}, emulating vector operations using scalar register files and execution units. Such architectures are only acceptable in cases with very low performance requirements.

A common approach to minimize the footprint of a dedicated VRF
is to employ narrow vector registers (storing fewer data elements) and reduce the number of vector execution lanes. However, this severely limits data parallelism, negatively impacting performance. This work presents an alternative strategy to reduce VRF costs without compromising vector width or parallelism.

\begin{figure}[t]
\centering
\includegraphics[width=0.9\columnwidth]{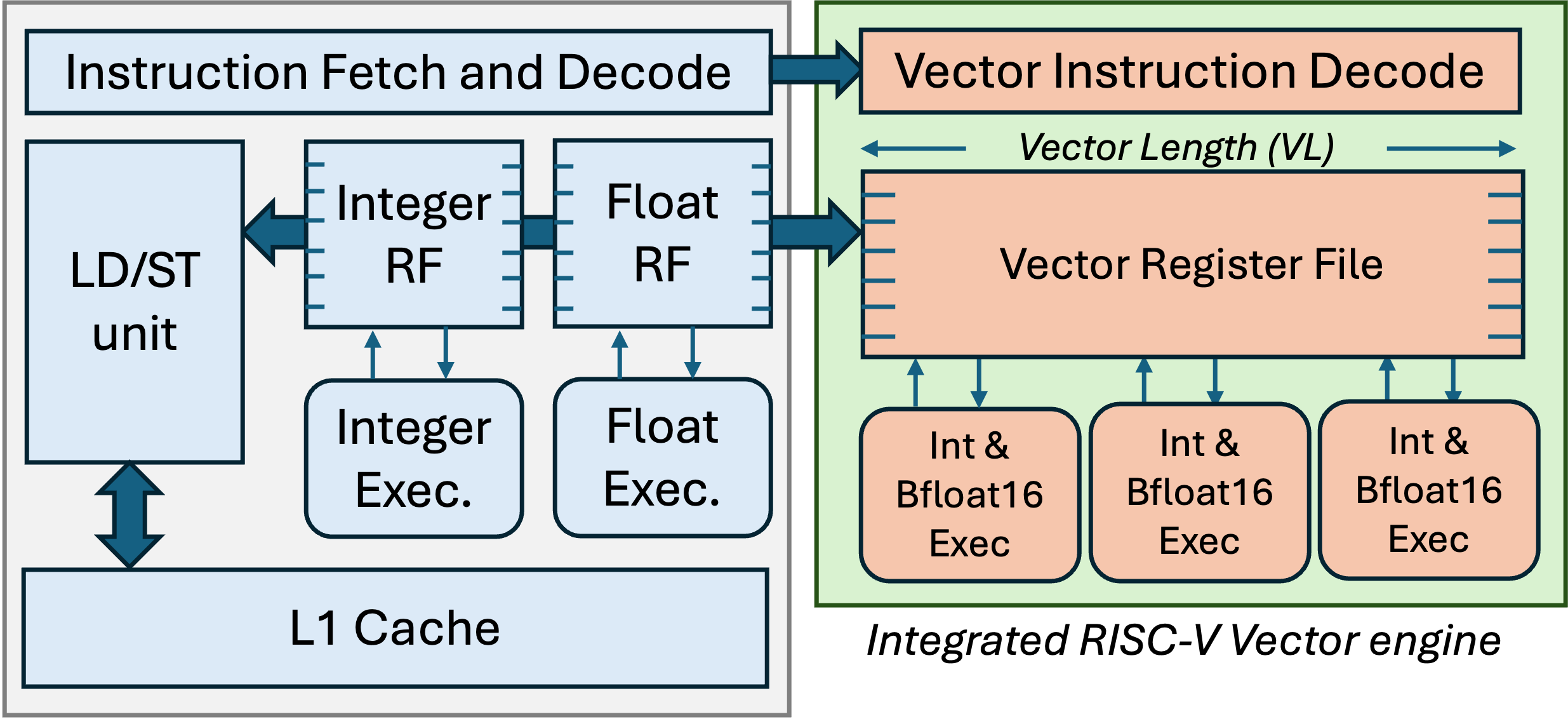}
\caption{The organization of a scalar RISC-V CPU augmented with an integrated VPU. The scalar and vector datapaths share the same load/store memory port to the L1 cache. }
\label{f:cpu_vpu_organization}
\end{figure}

Said approach is founded on the insight that most computation kernels in ML applications tend to use only a small subset of the architectural vector registers at any given time. This observation indicates that a VRF with \textit{fewer} physical registers than the total number dictated by the ISA would be adequate for most applications to still achieve full performance. This argument is supported also by a recent study for vector DSP kernels~\cite{redused_num_vreg}. Leveraging this insight, we hereby propose the use of a \textit{compact} VRF  (cVRF) that stores only a small subset of the architectural vector registers. The cVRF functions as a \textit{cache} by storing only the most recently accessed vector registers. The values of the remaining architectural registers are ``dispersed'' within the memory sub-system -- across the cache levels and main memory -- and loaded into the cVRF on-demand. The use of a cVRF does not impact ISA compliance; all architectural vector registers can still be used by the applications. What changes is their \textit{location} at any given time. While the overarching concept of \textit{register caching} has been explored within the contexts of superscalar processors~\cite{sup_cache} and Graphics Processing Units (GPU)~\cite{gpu_cache}, this is the first work -- to the best of our knowledge -- that implements such a scheme within a \textit{vector processing} environment. 
Overall, the contributions of this work can be summarized as follows:

\begin{itemize}
\item The introduction of Register Dispersion, a new process of dynamic management of the architectural vector registers. Register Dispersion uses a set of physical registers that is only a fraction of the set of architectural registers to markedly reduce the size of the VRF with no (or minimal) impact on application performance.
\item To facilitate Register Dispersion, a compact VRF design is proposed, which is much smaller (in terms of vector registers) than the ISA-mandated VRF size. The cVRF operates like a cache by storing only the recently accessed architectural registers, thereby making them readily available to the vector units. Binary code compatibility with the standard RVV ISA extension is fully maintained.
\item Extensive experimental evaluation and hardware cost analysis -- based on fully implemented and synthesized hardware implementations -- demonstrate the efficacy and efficiency of the proposed approach. The use of a cVRF yields significant VPU area savings of 53\%, as compared to using a full-size VRF, and 23\% savings in the \textit{total} CPU+VPU area. The total CPU+VPU power savings are 10\%, on average. Such hardware cost savings are achieved with no, or minimal, impact on performance. 
\end{itemize}
The presented architecture strikes a very effective cost-performance trade-off when adding vector processing capabilities to ultra low-cost processors. Consequently, the hardware acceleration of ML applications at the edge becomes more practical and feasible.
\section{Typical Low-Cost Processors with Integrated Vector Processing Units}
\label{sec:Low cost processor}

To better understand the critical importance of the hardware area and power cost in very resource-constrained environments, we first describe the typical organization of a processor core encountered in such systems.

\subsection{Architectural Considerations}
Ultra low-cost processors prioritize area efficiency and power consumption over peak performance. To achieve this goal, they typically adopt a simplified core design with trade-offs in certain architectural features. A typical organization for a low-cost processor features a lightweight single-issue datapath with no out-of-order capabilities. Additionally, the instruction and data caches are usually smaller in size and have lower associativity, as compared to high-performance processors.

In this vein of low-area prioritization over peak performance, the various design decisions also impact the number of pipeline stages. To minimize the overall chip area, these processors employ a reduced clock frequency, which allows for simpler and smaller circuit elements. In conjunction with lower clock speeds, implementing fewer pipeline stages can further contribute to area savings. 

Nevertheless, the insatiable demand for ML/AI applications, even at the edge, puts extra strain on the computational capabilities of ultra low-cost processors. They are simply not powerful enough -- especially in terms of throughput -- to handle these emerging applications to a level that makes the user experience satisfactory. Hence, to enhance the computational capabilities of ultra low-cost processors, one potential approach is to integrate a VPU. %within the execution stage of the pipeline. 
The VPU can be designed to handle vector operations efficiently, allowing for the parallel processing of multiple data elements. It is accessed using special vector instructions, as part of an ISA extension. These instructions require -- by ISA definition -- a vector register file. The VRF should be integrated alongside the existing scalar register file of the processor. The organization of an ultra low-cost processor augmented with an \textit{integrated} VPU is illustrated in Fig.~\ref{f:cpu_vpu_organization}. In contrast to \textit{integrated} VPUs, there are also \textit{decoupled} VPU architectures~\cite{Riscv2,vicuna,bigvlittle,ara2}, whereby the VPU has a dedicated path to the data cache.

The integrated VPU shares the fetch and decode pipeline stages with the scalar core to optimize resource utilization. Vector processing relies on a number of parallel execution units to effectively extract data-level parallelism. These execution units are directly fed by the vector registers of the VRF. The vector \textit{length} is a crucial parameter that denotes the bit-width of each vector register and determines the maximum degree of parallelism that can be exploited. Loading and storing these long vectors from/to main memory puts pressure on the cache/memory sub-system. To contain this pressure, the selected vector length should \textit{not} exceed the cacheline size; i.e., the maximum supported vector length is \textit{bounded} by the size of the cacheline. This constraint ensures that each vector load instruction can be served with a single micro-operation.

\subsection{Area Overhead of Integrating a VPU into a Low-Cost CPU}
\begin{figure}[h]
    \centering
    \includegraphics[width=0.85\columnwidth]{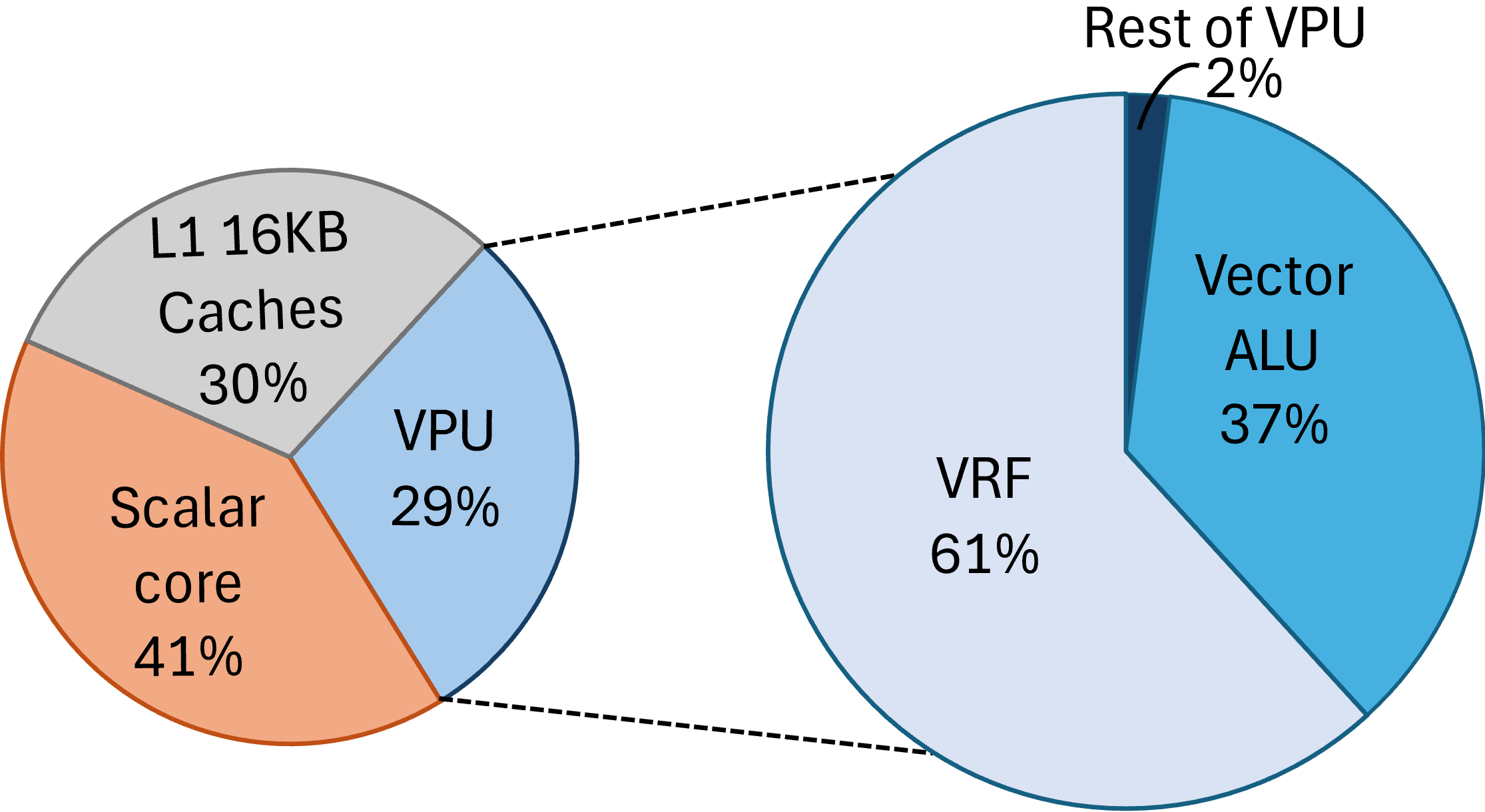}
    \caption{Percentage breakdown of the hardware area consumption of the main components of a RISC-V CPU with an integrated VPU. The VRF consumes 61\% of the VPU area.}
    \label{f:vrf_area}
\end{figure}

Adding a VPU to a low-cost CPU adds a significant amount of extra hardware area. To evaluate this overhead, a RISC-V compliant vector unit was integrated within the Codasip L31 RISC-V core~\cite{l31} with an organization similar to Fig.~\ref{f:cpu_vpu_organization}, and implemented at 28 nm technology. The VPU uses a vector length of 256 bits and has 8 vector lanes with 32-bit ALUs supporting 8-, 16-, or 32-bit integer and \texttt{bfloat16} calculations within the vector execution units. The harware area contribution of each component is shown in Fig.~\ref{f:vrf_area}. The L1 cache area refers to the cost of the storage memory, since the cache controller is built into the scalar core. As can be seen, the VRF consumes more than half (61\%) of the VPU's total area. While a reduction in the vector \textit{length} would yield substantial area savings, it would negatively affect the performance, since it would decrease the exploitable data-level parallelism.

This work demonstrates that a more effective approach in lowering the hardware area cost of the VRF with no/minimal effect on performance is to decrease the ``height'' of the VRF, i.e., the number of vector registers that the VRF can store.

\section{Register Dispersion Using a Compact Vector Register File}
\label{sec:register cache organization}

The proposed Register Dispersion technique dynamically manages the architectural vector registers to ensure that the recently accessed ones are readily available and close to the VPU. Instead, unused and not-recently-used architectural registers are kept within the memory sub-system; i.e., the cache and/or main memory.

\subsection{Microarchitecture of a Compact VRF}

The recently accessed vector registers are stored in a compact VRF (the cVRF), which is much smaller -- in terms of the number of vector registers that it can store -- than a full-size VRF that can store all architectural vector registers. The cVRF operates like a traditional \textit{cache}. Hence, in addition to the register \textit{data}, some additional meta-data are also stored to serve as \textit{tags}. The tag of a cVRF entry is used to specify the \textit{architectural} vector register that is currently ``mapped'' to this particular \textit{physical} register entry. Since the cVRF is, by definition, compact, it is implemented as a fully-associative cache to maximize its utilization without incurring undue hardware overhead. Thus, an architectural vector register can be stored anywhere within the cVRF.

Without loss of generality, we target the RISV-V ISA in the proposed implementation of the Register Dispersion mechanism. The RISC-V ISA allows the vector instructions to use masking. The vector mask is located exclusively in architectural vector \texttt{v0}, thereby making it advantageous to \textit{always} maintain this register within proximity of the VPU. Hence, architectural register \texttt{v0} is always stored in an additional, dedicated register, and it is never dispersed into the memory sub-system. Register Dispersion and the cVRF handle the management of the \textit{remaining} 31 registers.

All 31 \textit{architectural} vector registers have their own specially reserved address in memory. In other words, there is a fixed region in memory that is reserved for the 31 ISA-defined vector registers. At any given time, if a register is not located within the cVRF, it can be found at its dedicated pre-assigned memory address within the cache/memory sub-system.

\begin{figure}
    \centering
    \includegraphics[width=0.98\columnwidth]{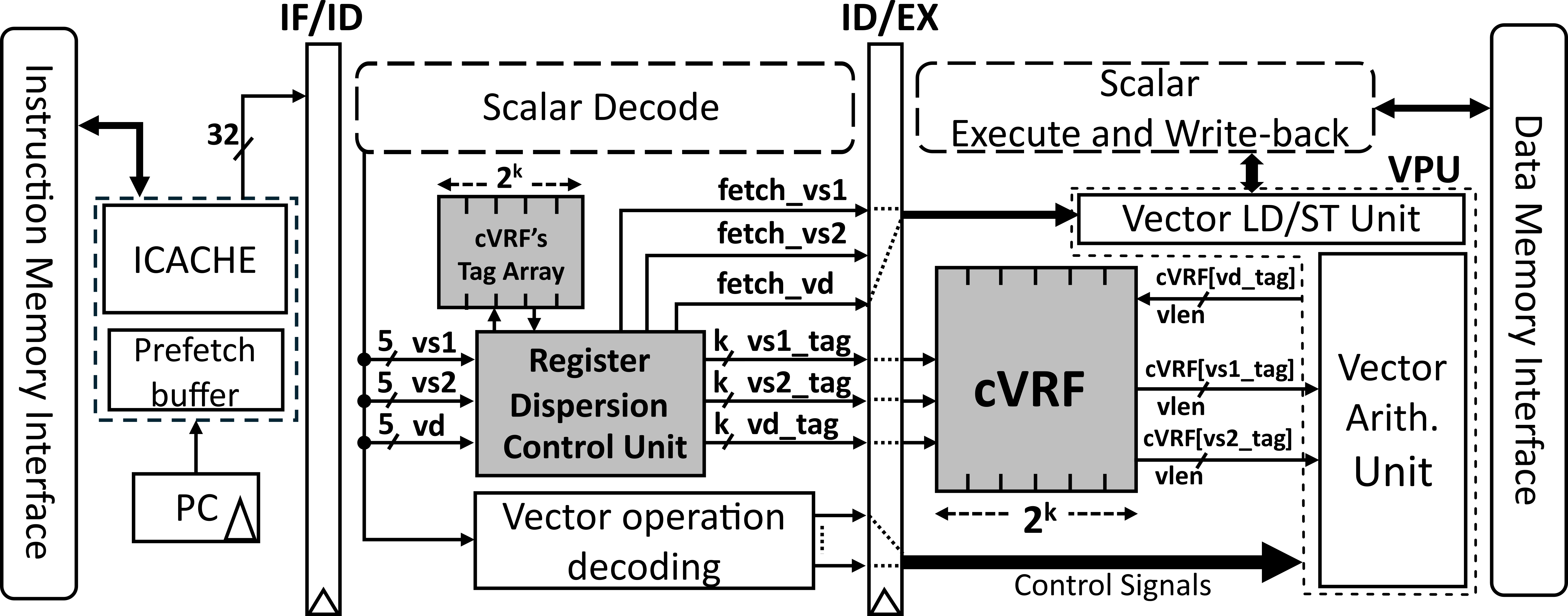}
    \caption{The micro-architecture of the proposed Register Dispersion technique. All components comprising the new mechanism are highlighted in grey. The tags in the cVRF's Tag Array specify the architectural vector registers that are currently ``mapped'' to physical register entries within the cVRF.}
    \label{f:cVRF}
\end{figure}

The micro-architecture of the proposed mechanism -- and its location within the processor pipeline -- is illustrated in Fig.~\ref{f:cVRF}, with all the Register Dispersion components highlighted in grey. As can be seen, the cVRF is, in fact, disaggregated: its \textit{tag} array is physically located in the Instruction Decode (ID) pipeline stage, while the actual \textit{data} registers (i.e., the physical vector registers) are located within the Execute (EX) stage. As previously mentioned, the cVRF's Tag Array contains the mappings of the architectural vector registers to the physical registers within the cVRF. A lack of a tag (mapping) indicates that the corresponding slot within the cVRF is free/available.

The cVRF's vector registers (located in the EX stage) are organized as a circular FIFO queue, with head and tail pointers. The head pointer always points to the register with the longest residency in the cVRF, whereas the tail pointer points to the next free slot.

\subsection{Dispersion Mechanism}

\subsubsection{Determining vector operand location}
Once a vector instruction enters the ID stage, its two source operands \textit{and} its destination operand must be read from the cVRF. Note that the destination operand is also needed, because the RISC-V ISA has numerous instructions that utilize the destination register in their execution (e.g., \texttt{vmacc}, \texttt{vmadd}, etc.). Thus, the Register Dispersion mechanism first checks to see which (if any) of these three operands are currently located within the cVRF. This is achieved by accessing the Tag Array independently for each of the three operands. The process of checking for cVRF residency is performed serially for all three operands to accommodate for the cases where cVRF evictions are required, as will be explained shortly.

If there is a hit in the Tag Array, the operand is already within the cVRF, and the tag hit location indicates the corresponding location within the cVRF (i.e., the physical mapping). This physical location (index) is carried -- through the ID/EX pipeline register -- to the next pipeline stage, where the cVRF will be accessed.

On the other hand, if there is a miss in the Tag Array, then the missing operand must be fetched from the memory sub-system. The mechanism's Control Unit stalls the processor and proceeds differently, based on cVRF slot availability.

\subsubsection{Replacement policy}
If there is at least one free slot in the cVRF, then the control unit dynamically generates appropriate \texttt{load} micro-operations to load the operand from memory into the cVRF, in the location pointed to by the circular FIFO's tail pointer. Recall that the memory address of each architectural register is a priori fixed and known. 
Again, the physical location/index in the cVRF is carried to the next pipeline stage.

If there are no free slots within the cVRF, then an \textit{eviction} must first be performed, before loading the missing operand. In its current incarnation, the proposed mechanism employs a FIFO replacement policy, due to its simplicity. By leveraging the existing circular FIFO implementation of the cVRF, a FIFO replacement policy is achieved by simply evicting the register that is located at the head pointer of the cVRF. On the contrary, a Least Recently Used (LRU) policy would require additional counters. Despite its simplicity, the FIFO replacement policy will be demonstrated to achieve very high hit rates. After locating the evictee, the Control Unit first generates appropriate \texttt{store} micro-operations to store the evicted register to memory. The memory address is determined by the evictee's tag entry, i.e., its architectural register number. Subsequently, the Control Unit generates \texttt{load} micro-operations to load the missing operand from memory into the cVRF (in the location pointed to by the tail pointer). Same as before, this cVRF index is carried to the next pipeline stage.

Upon completion of the above-mentioned process, all three operands required by the vector instruction are located within the cVRF, and their indexes are stored within the ID/EX pipeline register. The vector instruction then proceeds to the EX pipeline stage. During this stage, the cVRF is accessed using the indexes from the previous (ID) stage, and the register values are sent to the VPU for instruction execution.

In summary, the Register Dispersion mechanism employs a simple Control Unit and a split-type cVRF -- with its Tag Array in the ID stage and its Data registers in the EX stage -- to dynamically manage the usage of the vector registers.
\begin{table}[t!]
\centering
\caption{Processor configuration}
\label{t:cpu_parameters}
\renewcommand\tabularxcolumn[1]{m{#1}}
\begin{tabularx}{1\columnwidth}{cX}
\hline
Scalar core & 
\begin{itemize}[leftmargin=15pt]
\item 
Codasip L31 RISC-V embedded CPU (RV32IMFCB)~\cite{l31}
\item 
Single-issue, 3 pipeline stages, 200 MHz clock frequency
\item
32 32-bit scalar and 32 32-bit floating-point registers
\item
L1I cache: 1-cycle hit latency, 2-way, 16 KB
\item 
L1D cache: 1-cycle hit latency, 2-way, 16 KB
\item 
Main memory: 1--5-cycle latency @200MHz, 2 MB
\end{itemize}\\
\hline
VPU & \begin{itemize}[leftmargin=15pt]
\item 256-bit vector engine with 8-lane configuration (32-bit elements $\times$ 8 execution lanes)
\item 
 Full VRF (32-Vregs) / cVRF (3/4/5/6/7/8/16-Vregs)
\end{itemize} \\
\hline
\end{tabularx}
\end{table}

\section{Experimental Evaluation}
\label{s:eval}

The experimental evaluation has a double-faceted objective: (a) demonstrate that a reduction in the number of registers stored within the vector register file has no effect on application performance; (b) show that the use of the proposed light-weight cVRF reaps substantial hardware area/power savings, as compared to a full-size VRF that stores all the ISA-defined architectural vector registers.

\subsection{Experimental Setup}

The commercially-available L31 low-cost 3-stage RISC-V microprocessor~\cite{l31} is augmented with a VPU and the proposed Register Dispersion mechanism were fully implemented in CodAL, an architecture description language, which facilitates the design of processors. The micro-architectural parameters of the processor are summarized in Table~\ref{t:cpu_parameters}.

For the \textit{performance} evaluations, the Codasip Studio simulator was used, which can \textit{cycle-accurately} simulate processors designed in the CodAL language while executing benchmark applications, and all the performance metrics were derived using this simulator. The applications employed in the experimental results and their characteristics are depicted in Table~\ref{t:apps}. 
The applications examined include benchmarks taken from the RiVEC benchmark suite~\cite{rivec} and Ara's application collection~\cite{ara2}. 
Machine learning kernels from Natural Language Processing (NLP), Convolutional Neural Networks (CNN), and transformers employed in Large Language Models (LLM) are also included. More specifically, the `DenseNet121 Layer105' and `ResNet50 Layer112' benchmarks correspond to layer 105 of DenseNet121~\cite{densenet} and layer 10 of ResNet50~\cite{resnet}, respectively. The execution of those CNN layers is mapped to an equivalent matrix multiplication after employing the im2col transformation on the corresponding input features and weights ~\cite{im2col}. Additionally, a self-attention layer of the Bert application~\cite{bert} -- implemented in a vectorized form using the FlashAttention-2 technique~\cite{flashattention2} -- is also included.

\begin{table}
\caption{Applications characteristics}
\label{t:apps}
\begin{tabular}{c|c|c}
\hline
Application & Domain & Characteristics \\
\hline
\hline
\multicolumn{1}{c|}{\multirow{2}{6.5em}{PathFinder}}&\multicolumn{1}{c|}{\multirow{2}{6.5em}{Grid Traversal}}&\multicolumn{1}{c}{Rows:32}\\
&&\multicolumn{1}{c}{Columns:32}\\
\hline
\multicolumn{1}{c|}{\multirow{2}{6.5em}{Jacobi-2D}}&\multicolumn{1}{c|}{\multirow{2}{5.5em}{Engineering}}& \multicolumn{1}{c}{Problem size:128}\\
&&\multicolumn{1}{c}{steps:10}\\
\hline
\multicolumn{1}{c|}{\multirow{2}{6.5em}{Somier}}&\multicolumn{1}{c|}{\multirow{1}{3em}{Physics}}& \multicolumn{1}{c}{Problem size:32}\\
&\multicolumn{1}{c|}{Simulation}&\multicolumn{1}{c}{steps:2}\\
\hline
\multicolumn{1}{c|}{\multirow{1}{6.5em}{GemV}}&\multicolumn{1}{c|}{NLP}&\multicolumn{1}{c}{(512  x  512) x 512}
\\
\hline
\multicolumn{1}{c|}{\multirow{2}{6.5em}{Dropout}}&\multicolumn{1}{c|}{\multirow{2}{1.5em}{ML}}& \multicolumn{1}{c}{Vector Length:131072}\\
&&\multicolumn{1}{c}{Scale:0.5}\\
\hline
\multicolumn{1}{c|}{\multirow{2}{6.5em}{fconv2d-7x7}}&\multicolumn{1}{c|}{\multirow{2}{2.5em}{CNN}}& \multicolumn{1}{c}{256 x 256}\\
&&\multicolumn{1}{c}{filter size:7}\\
\hline
\multicolumn{1}{c|}{\multirow{1}{6.5em}{DenseNet121}}&\multicolumn{1}{c|}{\multirow{2}{2.5em}{CNN}}& \multicolumn{1}{c}{\multirow{2}{9em}{(32 x 1152)x(1152 x 64)}}\\
\multicolumn{1}{c|}{\multirow{1}{6.5em}{Layer105}}&&\\
\hline
\multicolumn{1}{c|}{\multirow{1}{6.5em}{ResNet50}}&\multicolumn{1}{c|}{\multirow{2}{2.5em}{CNN}}& \multicolumn{1}{c}{\multirow{2}{9em}{(128 x 256)x(256 x 784)}}\\
\multicolumn{1}{c|}{\multirow{1}{6.5em}{Layer10}}&&\\
\hline
\multicolumn{1}{c|}{\multirow{4}{7em}{FlashAttention-2}}&\multicolumn{1}{c|}{\multirow{4}{5.5em}{Transformer}}& \multicolumn{1}{c}{Seq. Length:200}\\
&&\multicolumn{1}{c}{Hidden Dim.:64}\\
&&\multicolumn{1}{c}{Block size row:1}\\
&&\multicolumn{1}{c}{Block size column:128 }\\
\hline
\end{tabular}
\end{table}

For the \textit{hardware cost} analysis, the RTL implementation of the low-cost processor with its VPU and the Register Dispersion technique were generated using Codasip Studio. The RTL was then synthesized using the Cadence digital implementation flow and a 28 nm standard-cell library for the hardware cost evaluation.

\subsection{Performance results}
\label{ss:vectorized}

\subsubsection{Speedup achieved with a full-sized VRF}
The addition of a VPU to the scalar CPU core can reap significant performance uplifts when executing workloads with high degree of DLP. The second column of Table~\ref{t:perf_with_full_vrf} shows the performance speedups achieved in the examined vectorized benchmark applications when using a VPU, as compared to scalar (i.e., CPU-only) execution. In this experiment, the VPU uses a full-size VRF that can store all 32 architectural vector registers, and the vector length is set to 256 bits to align with the common cache line size of 32 bytes used in many low-cost processors~\cite{cortexm55,Espressif}. As can be seen, the speedups range from around 4.3$\times$ to 8$\times$, thereby corroborating the benefit of augmenting the CPU with a VPU.

\begin{table}[h]
\centering
\caption{Speedups achieved when using a VPU with a full-size VRF and 256-bit vector length over scalar execution. The VRF utilization percentages are also shown.}
\label{t:perf_with_full_vrf}
\begin{tabular}{c||c||c|c}
\hline
Applications & Speedup & \#Active Vector  & VRF \\
             &        &     Registers     &util. (\%) \\
\hline
PathFinder& 7.99$\times$ & 6 & 18\% \\
Jacobi-2D& 6.48$\times$ & 7 & 21\% \\ 
Somier&7.82$\times$&14&44\%\\
GemV&6.89$\times$&9&28\%\\
Dropout&4.3$\times$&3&9\%\\
fconv2d-7x7&7.74$\times$&15&47\%\\
DenseNet121 Layer105& 7.82$\times$ & 4 & 12\% \\
ResNet50 Layer10& 7.63$\times$ & 4 & 12\% \\
FlashAttention-2&7.91$\times$&32&100\%\\
\hline
\end{tabular}
\end{table}

Nevertheless, this dramatic improvement in performance comes at a significant hardware cost. As shown in Fig.~\ref{f:vrf_area}, the addition of a VPU -- with its full-size VRF -- consumes area comparable to the area occupied by the scalar core. Such area cost is often prohibitive for ultra low-cost processors that are resource-constrained. To decrease the area overhead, we focus on the VRF of the VPU, which occupies 61\% of the entire VPU area (see Fig.~\ref{f:vrf_area}).

\subsubsection{Reducing the number of vector registes does not hurt performance}
The founding premise of the proposed Register Dispersion mechanism is the observation that most applications tend to use very few architectural registers during execution. This behavior is obvious in the third column of Table~\ref{t:perf_with_full_vrf}, which lists the number of architectural vector registers used by the evaluated benchmark applications. As can be seen, in most cases, the number
of used registers is much lower than the 32 defined by the ISA. The use of very few vector registers is also reflected in the VRF utilization percentages shown in the fourth column of Table~\ref{t:perf_with_full_vrf}.
Special reference should be made to the `FlashAttention-2' benchmark. Even though said application uses all 32 vector registers throughout its entire execution, it only operates on very few registers at any given time, as will be demonstrated later. In other words, each execution phase works on a small subset of the 32 registers.

Given this salient attribute shared by most of the examined applications, it is worthwhile to investigate how many registers a cut-down VRF -- i.e., the proposed cVRF -- must be able to store to reach the same performance levels as when using a full-size VRF.

\begin{figure}[t!]
\centering
\begin{minipage}{0.8\columnwidth}
    \centering
    \includegraphics[width=\textwidth]{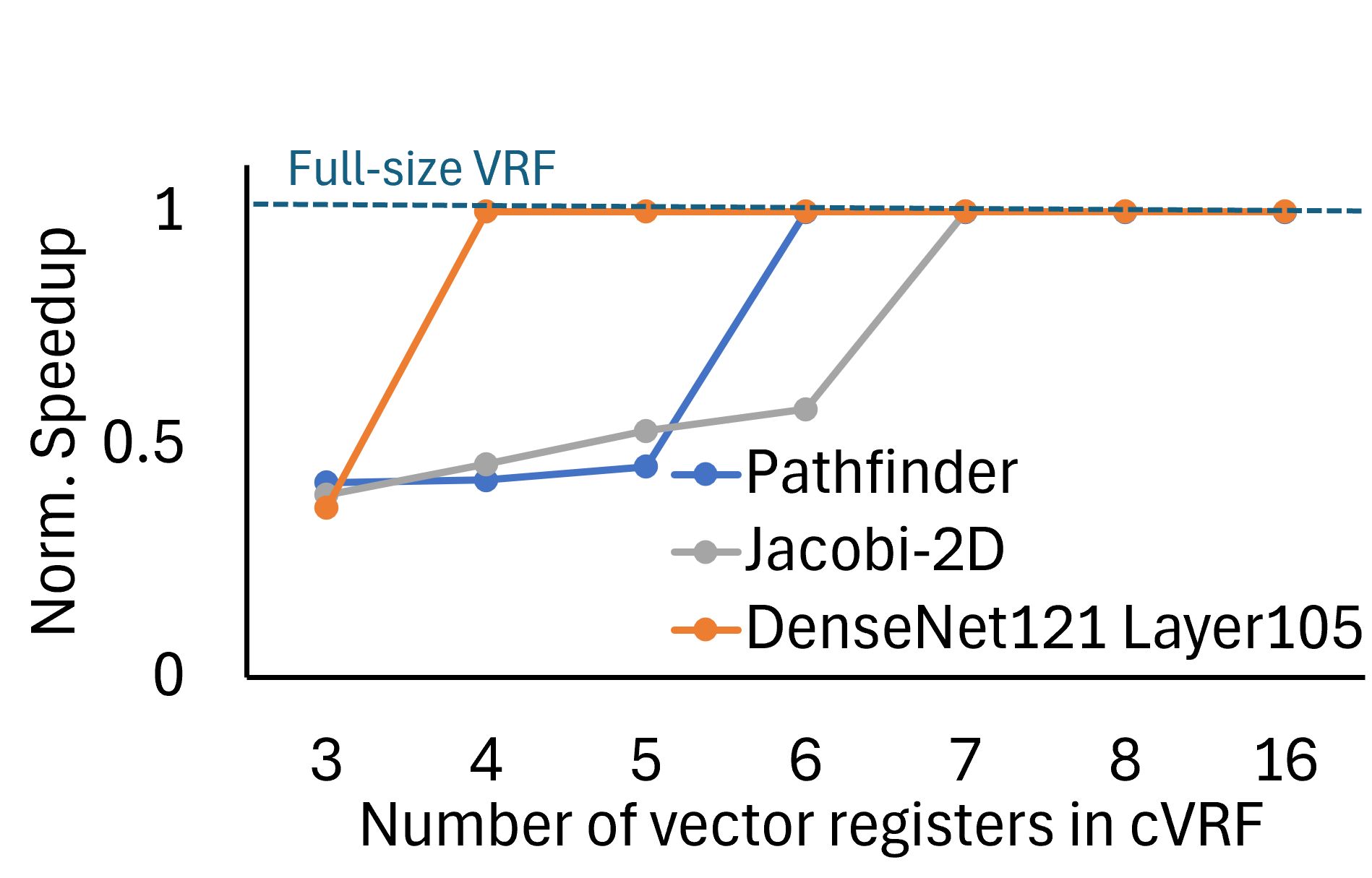}
    {\small (a) Achieved performance}
\end{minipage}%
\hfill
\begin{minipage}{0.8\columnwidth}
    \centering
    \includegraphics[width=\textwidth]{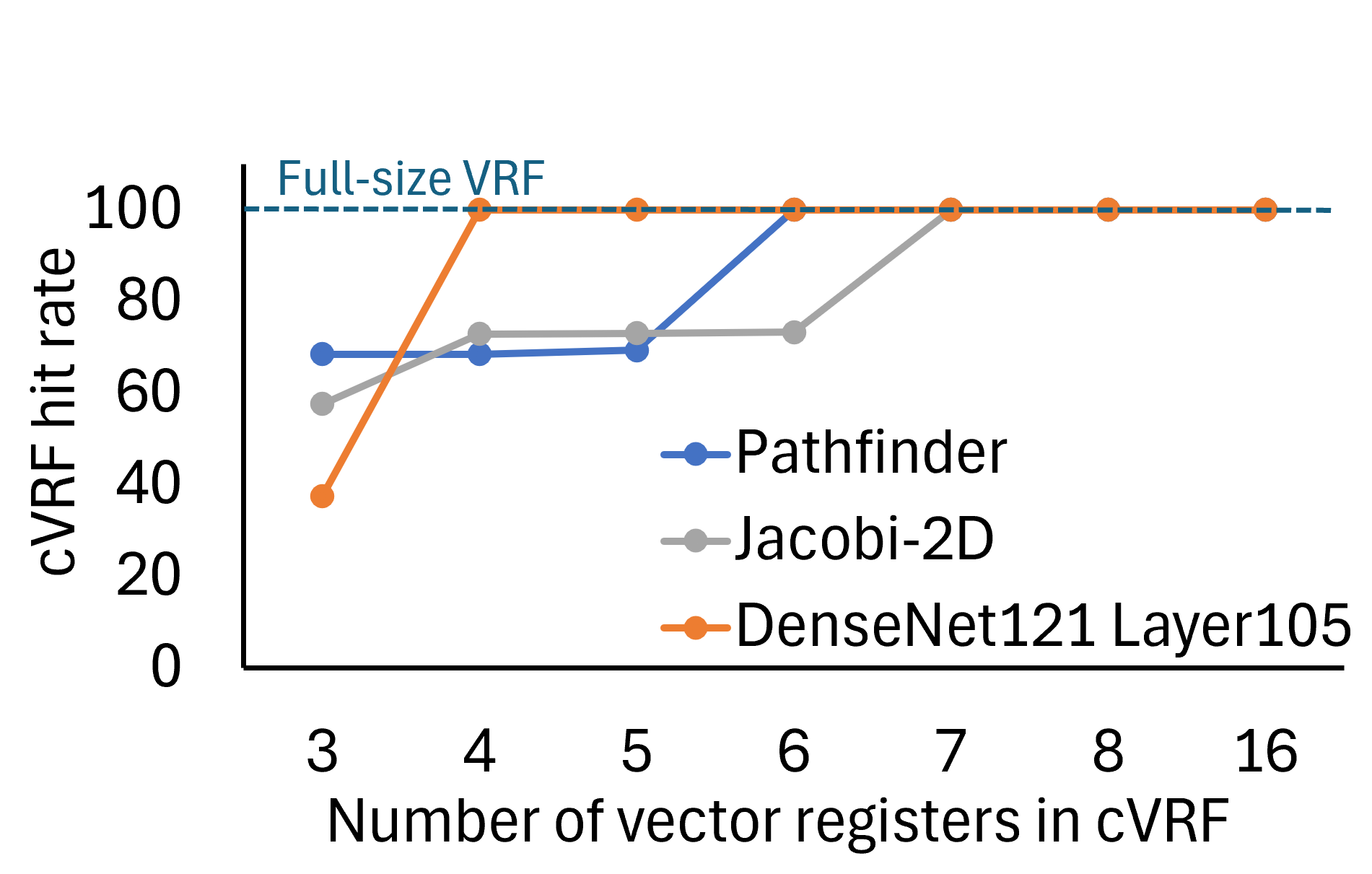}
    {\small (b) Hit rates}
\end{minipage}
\caption{(a) The performance achieved by various cVRF sizes, and (b) the hit rates in cVRFs of various sizes, when executing three indicative benchmark applications. All results are normalized to the performance of a full-size VRF.}
\label{f:perf_cvrf_sizes}
\end{figure}

\begin{figure}
    \centering
    \includegraphics[width=0.99\columnwidth]{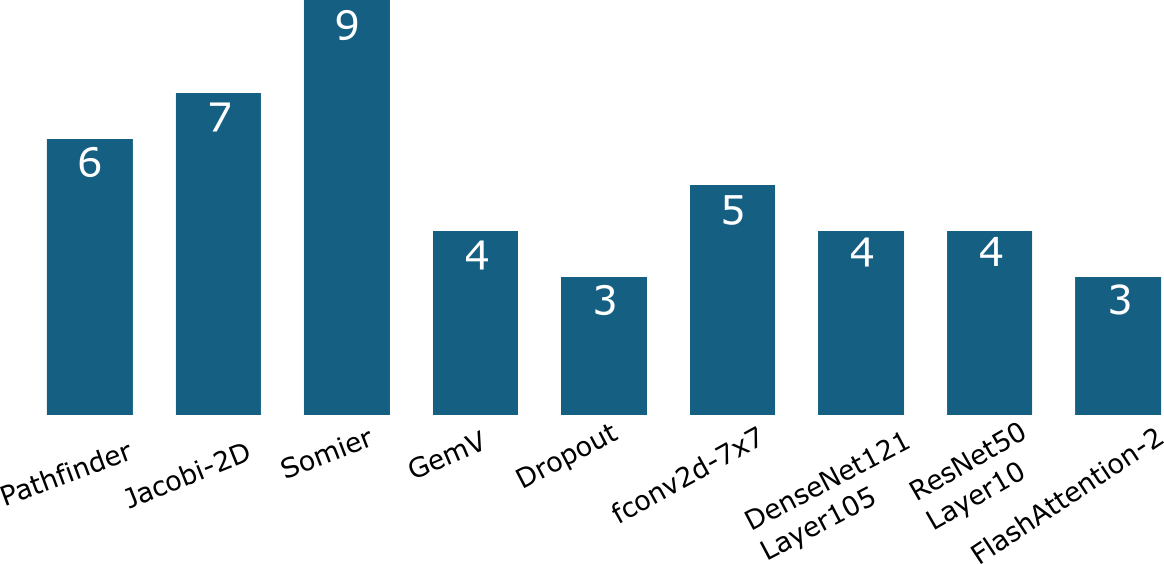}
\caption{The minimum number of vector registers required to achieve a hit rate above 95\% in the compact VRF.}
\label{f:95_hitratio}
\end{figure}

Toward this end, and based on the numbers of used vector registers reported in Table~\ref{t:perf_with_full_vrf}, we investigate the use of cVRFs of various sizes. Specifically, we investigate cVRFs that can store from 3 to 16 vector registers. To aid visual clarity (and since all applications show similar trends), the performance results for an indicative subset of the examined benchmarks are illustrated in Fig.~\ref{f:perf_cvrf_sizes}(a). All results are normalized to the performance achieved when using a full-size VRF that can store all 32 architectural vector registers defined by the ISA. This normalization reference is the best possible performance that can be achieved. The vector length is set to 256 bits in all configurations. As shown in Fig.~\ref{f:perf_cvrf_sizes}(a), the performance of the smaller cVRFs is markedly worse than the full-size VRF, because they cannot fit all the registers accessed by the applications. However, as the size of the cVRF increases, the performance improves and eventually -- when the size of the cVRF is 8 -- reaches the performance of the full-size VRF. Hence, this experiment demonstrates that \textit{storing 8 vector registers} in the cVRF is \textit{sufficient to achieve full performance} (i.e., no degradation) \textit{for a wide range of vector applications}.

Fig.~\ref{f:perf_cvrf_sizes}(b) shows the hit rates in the various examined cVRFs for the same indicative subset of benchmarks. As the number of stored vector registers approaches the maximum number used by each application, the hit rates increase. As is evident, the bigger cVRFs (of sizes 7 and above) have almost 100\% hit rate. On the other hand, the small cVRFs of sizes 3 to 5 suffer from many misses in the `Pathfinder' benchmark, which uses more registers that can fit within the cVRF. Similarly, the cVRFs of sizes 3 to 6 exhibit many misses in the `Jacobi-2D' application, for the same reason. 

\begin{figure*}[t!]
    \centering
    \includegraphics[width=1.85\columnwidth]{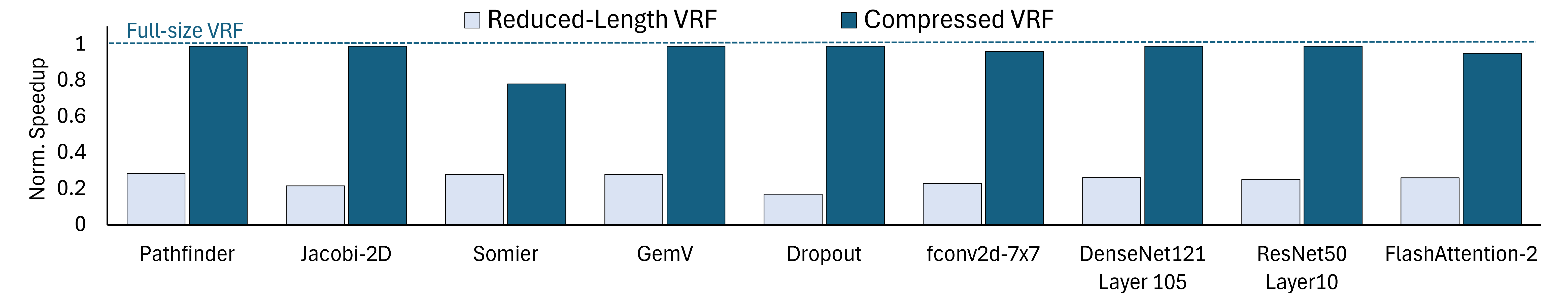}
    \caption{Performance comparison between Register Dispersion with a cVRF of 8 256-bit vectors versus an \textit{equal-area} VRF with 32 vectors of reduced 64-bit length. The results are normalized to the performance achieved when using a VPU with a full-size VRF (i.e., 32 vectors of 256-bit length).}
    \label{t:perf_win}
\end{figure*}

To further support our assertion that storing eight vector registers within the cVRF is sufficient in achieving peak performance, we determined the minimum number of registers required to achieve a hit rate exceeding 95\% within the cVRF for all examined applications. The results are illustrated in Fig.~\ref{f:95_hitratio}. Similar to the previously mentioned behavior of the `FlashAttention-2' benchmark, even though some applications require more than eight vector registers throughout their entire execution, not all the vector registers are utilized \textit{simultaneously} at any given time. This contributes positively to the hit rate within the cVRF. This is precisely the reason why the `FlashAttention-2' application exhibits a very high hit rate even with a cVRF of only 3 vector registers, even though it uses all 32 vector registers over its entire execution. Again, the results in Fig.~\ref{f:95_hitratio} demonstrate that a cVRF with eight registers is adequate for most cases.

\subsection{Comparison against a 32-register VRF with reduced vector length}

As explained in Section~\ref{sec:Low cost processor}, another approach to reducing the hardware area of the VRF is to decrease the vector \textit{length} of each vector in the VRF, instead of the number of stored registers. We hereby demonstrate that this alternative approach leads to much lower application performance, as compared to the performance reaped with Register Dispersion. For a fair comparison, we compare two different VRF implementations that have \textit{equal area}: (1) a cVRF that can store 8 vector registers, each of 256-bit vector length; %(denoted as `impl(256,8)')
and (2) a VRF that can store all 32 architectural vector registers, but with a reduced vector length of 64 bits. 
%(denoted as `impl(64,32)'). 
In other words, the first implementation corresponds to the proposed Register Dispersion approach of reducing the VRF ``height'' by a factor of 4, whereas the second implementation corresponds to the approach of reducing the ``width'' (i.e., vector length) of the full-size VRF by a factor of 4. 

The obtained results are depicted in Fig.~\ref{t:perf_win} and are normalized to the performance achieved when using a VPU with a full-size VRF (i.e., 32 vectors of 256-bit length). As can be seen, the Register Dispersion setup achieves near-identical performance to the system with a full-size VRF in \textit{most} benchmark applications. With the `Somier' and `fconv2d' benchmarks, the performance is lower, since these applications access more than 8 architectural vector registers. Nevertheless, the performance of `fconv2d' is not affected considerably. This can be attributed to the strategic grouping and unrolling of vector register names within the application's code.

Interestingly, the performance results for all six benchmark applications from the ML-related domains (i.e., NLP, general ML, CNN, and Transformer) are excellent under Register Dispersion. As can be seen in the right six sets of bars in Fig.~\ref{t:perf_win}, the cVRF approach comes very close to matching the performance of a full-size VRF with only a quarter of its size. This behavior clearly indicates that the proposed mechanism is, indeed, a promising approach to accelerating ML applications on ultra-low-cost processors on the edge.

Compared to Register Dispersion, the configuration with reduced vector length performs much worse across all benchmarks. The reason for this huge disparity in achieved performance is the severely diminished data-level parallelism when decreasing the vector \textit{length}.

\begin{figure}[h!]
\centering
\includegraphics[width=0.9\columnwidth]{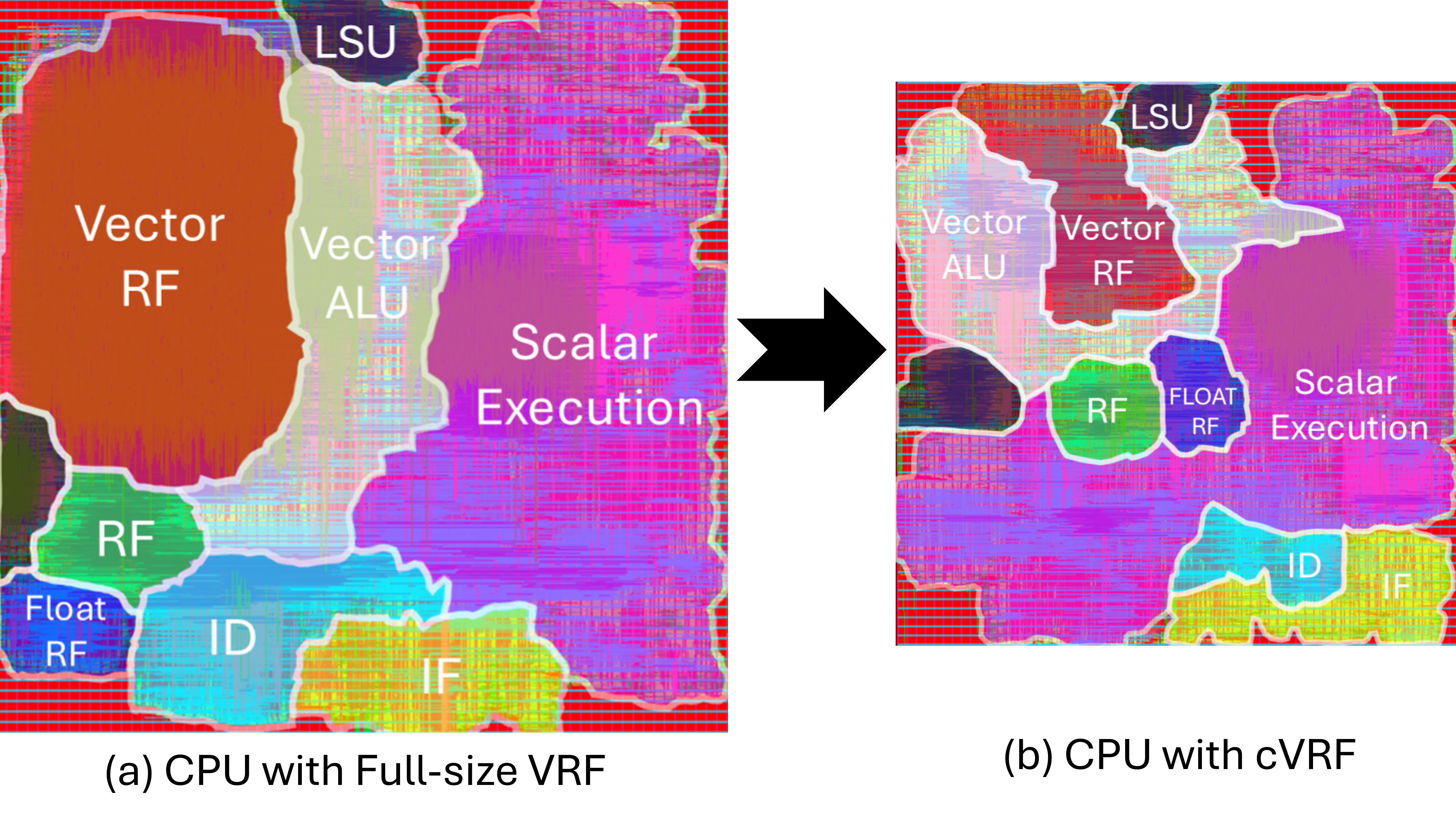}
\caption{The physical layout of the two versions of the examined CPU. Highlighted regions include the Instruction Fetch (IF) logic, the Instruction Decode (ID) logic, the Load-Store Unit (LSU) of the scalar core, the scalar execution units, and the integer and floating-point register files. The VPU consists of the regions of the VRF and the Vector ALU. Compacting the VRF also altered the relative placement of the CPU modules.}
\label{f:layout}
\end{figure}

\begin{figure*}[h!]
    \centering
    \includegraphics[width=1.85\columnwidth]{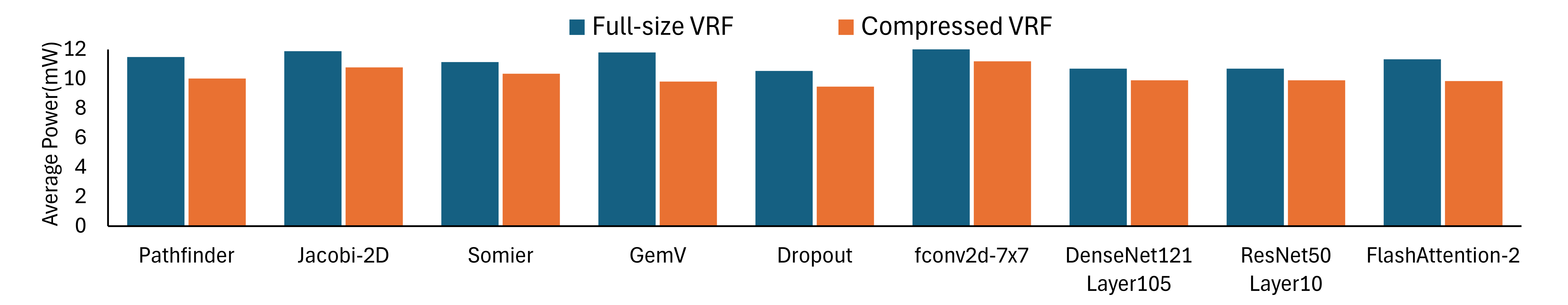}
    \caption{The average power consumption for all examined applications when using a CPU with a full-size VRF and when using the same CPU with the proposed cVRF.}
    \label{f:power_per_appl}
\end{figure*}

\subsection{Hardware cost analysis}
\label{ss:hardware_cost}

To assess the hardware area and power savings achieved by the Register Dispersion mechanism, we compare a CPU utilizing a compact Vector Register File (cVRF) capable of storing 8 vector registers with a vector length of 256 bits to an equivalent CPU employing a full-size VRF capable of storing all 32 vector registers. Both evaluated configurations operate at a clock frequency of 200~MHz, which falls within the typical range observed in low-cost processors~\cite{stm}. 

\subsubsection{Area Savings}
\label{sss:area_savings}

The physical layout of both implementations at 28 nm is illustrated in Fig.~\ref{f:layout}. The layout on the left corresponds to the CPU with a VPU that utilizes a full-size VRF, whereas the layout on the right side corresponds to the same CPU, but with its VPU designed using a cVRF. Both layouts focus only on the logic area of the CPU, excluding the L1 instruction and data caches. For both scalar and vector register files, we assume a flip-flop-based standard-cell-based implementation. Moving to latch-based designs, as done in~\cite{Spatz}, is possible and would equally benefit both the full-size VRF and the cVRF implementations.

Both layout photos highlight the main blocks of the scalar core and the vector engine. The scalar execution logic occupies the majority of the area, since it includes both an integer ALU with a 32-bit multiplier and divider and a complete single-precision floating point unit. The scalar CPU also includes two separate register files; one is used for integer operands, and the other for floating-point operands.

By reducing the number of stored registers from 32 to 8, the area occupied by the VRF decreases by 3.5$\times$. The reduction does not reach 4$\times$, because the proposed approach incurs the extra cost of the cVRF tag array and the Register Dispersion control unit. The reduction in the total VPU area (i.e., the VRF and the Vector ALU) is around 53\%. This translates to a 23\% area reduction in the \textit{total processor area} (i.e., \textit{entire} CPU+VPU setup).

The reported area savings are accompanied with a reduction in routing congestion, as a consequence of the VRF's wide read and write ports. The cVRF -- that includes only 8 vector registers -- allows the flip-flops of the register file to spread evenly in space, thereby reducing routing congestion. This attribute is better highlighted by the layout of the cVRF shown in Fig.~\ref{f:layout}(b), where the vector ALU is physically wrapped around the smaller cVRF. In contrast, when using a full-size VRF, the two modules are automatically placed next to each other, as shown in Fig.~\ref{f:layout}(a) to handle the routing of the 256-bit-wide data buses that connects them,

\subsubsection{Application-level average power savings}
Each examined application was executed on the RTL of both CPUs under comparison to estimate the corresponding switching activity. Non-ML applications were driven by uniformly random inputs, while ML applications were executed on relevant datasets. Subsequently, the average power consumption of each CPU while executing each application was estimated on the final extracted netlist after layout, including all interconnect parasitics. The comparison results of the per-application average power consumption are presented in Fig.~\ref{f:power_per_appl}. The CPU utilizing a cVRF with register dispersion exhibits average power savings of 10\%, as compared to the CPU with a full-size VRF.

In summary, the proposed mechanism is very effective in reducing the hardware cost of the VPU, as compared to using a full-size VRF. Most importantly, these hardware savings are achieved with no/minimal negative impact on performance.
\section{Related Work}
\label{s:related}

As previously mentioned, the work presented in this article is the first -- to the best of our knowledge -- to incorporate a reduced-size register file that operates like a cache within a \textit{vector} processor. Nevertheless, this elemental concept has also been exploited in other computing domains~\cite{sup_cache,gpu_cache}. The targeted domains so far have been superscalar processors~\cite{sup_cache} and Graphics Processing Units (GPU)~\cite{gpu_cache}. Specifically, the design proposed in~\cite{sup_cache} aims to reduce the latency and energy requirements of the register file of a superscalar core for similar IPC performance. By using a \textit{dedicated two-level} and banked register file organization, the proposed approach reduces the register-file size and the port requirements. On the contrary, our work uses a smaller, \textit{single-level} register file, and it relies on the \textit{existing} levels of the cache/memory sub-system to manage the spill-over registers.

The design in~\cite{gpu_cache} reduces the dynamic energy consumption of GPUs by utilizing energy-efficient small caches between the register file and the execution units. By using a hierarchical register-file architecture, certain registers are cached within existing operand buffering structures that are located closer to the execution pipelines of the GPU. Once again, this approach is founded on a two-level organization, as opposed to our work that uses all the levels of the existing cache/memory hierarchy.

Low-cost processors and vector processing units have been the subject of significant research, particularly in the domain of embedded systems and the Internet-of-Things (IoT)~\cite{low_cost_on_iot,minimal_vec_proc}, where resource constraints and energy efficiency are critical. The significance of the hardware cost of the VRF, in particular, was also recognized in~\cite{Spatz}. Specifically, the Spatz vector processor presented in~\cite{Spatz} demonstrates how compact vector units with minimal VRFs can deliver high efficiency while leveraging the RISC-V Vector (RVV) Extension. In contrast to the cVRF approach proposed in this paper, Spatz uses a full 32-vector VRF and lowers its hardware cost though the use of optimized latch-based Standard Cell Memory (SCM). With superior area and energy efficiency, the Spatz architecture is well-suited for edge and IoT applications. Nevertheless, this approach is seen as orthogonal to our technique, and -- as mentioned in Section~\ref{sss:area_savings} -- it could be used to further lower the hardware cost of the cVRF. The work in~\cite{integrvec} proposed the integration of vector units within a low-cost scalar processor, but no hardware optimizations were made pertaining to the VPU. To reduce the area and power overhead in vector processors, some researchers have proposed the elimination of the vector register file and the accessing of data directly from memory~\cite{Ahromma}. 

The Adaptable Vector Architecture (AVA)~\cite{AVA} targets micro-architectures designed for short decoupled vectors and facilitates the reconfiguration of VRF operations to also effectively handle \textit{long} vectors. This is achieved by introducing an additional register to store intermediate data. The AVA mechanism also implements a two-level register renaming scheme, which includes an intermediate mapping of architectural registers to Virtual Vector Registers (VVRs). However, these enhancements -- that aim to increase flexibility -- inevitably increase the area overhead, rendering AVA unsuitable for edge computing environments. This added overhead and complexity to the VPU cannot be justified within the context of resource-contrained, \textit{low-cost} processors, which is precisely the target of our work. 

Single-Instruction Multiple Data (SIMD) units are often preferred over vector processors, especially for embedded systems, due to their simpler design. SIMD processing has been used for multimedia applications on mobile devices~\cite{simd_multimedia}. The Sparrow architecture~\cite{sparrow} leverages the scalar integer register file for SIMD operations and uses a reduced number of SIMD instructions to achieve area savings.
\section{Conclusions}

The integration of a vector processing unit within the scalar core of a low-cost processor is a promising approach to significantly increase performance when executing data-level parallel applications. Given the increasing deployment of ML/AI applications at the edge, the ability to effectively exploit DLP in low-cost processors is of paramount importance. Nevertheless, the addition of a VPU entails the cost of the VRF, which is a substantial area consumer.

This work proposes the Register Dispersion mechanism, which builds on the observation that most ML applications use very few architectural vector registers. By employing a cut-down VRF that stores a small number of vector registers, Register Dispersion caches only the most recently accessed vector registers. The remaining registers are ``dispersed'' within the cache/memory sub-system. The proposed technique is demonstrated to yield significant area/power savings, as compared to using a full-size VRF. Most importantly, the hardware savings are reaped with no/minimal impact on performance. 

\section*{Acknowledgments}
This work was supported by a research grant from Codasip, a provider of customizable RISC-V IP and Codasip Studio design toolset, and its University Program to Democritus University of Thrace for ``RISCV vector processor design''.
\bibliographystyle{ACM-Reference-Format}
\bibliography{refs}

\end{document}